\documentclass[12pt,nofootinbib]{revtex4-2}
\usepackage{hyperref}
\usepackage{amsmath}
\usepackage{amsfonts}
\usepackage{graphicx}
\usepackage{multirow}
\usepackage{tikz}
\usepackage{ifthen}
\usepackage{siunitx}
\tikzset{>=latex}

\begin{document}
		\title{  Multi-Valued Quantum Neurons }
	\author{M.W. AlMasri} 

	\address{CyberSecurity and  Systems Research Unit, USIM, Malaysia}

	\begin{abstract}
	 The multiple-valued quantum logic is formulated systematically such that the truth values are represented naturally as unique roots of unity placed on the unit circle.  Consequently, multi-valued quantum neuron (MVQN) is based on the principles of multiple-valued threshold logic over the field of complex numbers.  The training of MVQN is reduced to the movement along the unit circle. A quantum neural network (QNN)  based on multi-valued quantum neurons can be constructed with complex weights, inputs, and  outputs encoded by roots of unity and an activation function that  maps the complex plane into the unit circle. Such neural networks enjoy fast convergence and higher functionalities compared with quantum neural networks based on binary input with the same number of neurons and layers. Our construction can be used in analyzing the energy spectrum of quantum systems. Possible practical applications can be found using the quantum neural networks built from   orbital angular momentum (OAM) of light  or multi-level systems such as  molecular spin qudits.  
	\end{abstract}
\maketitle
	\section{Introduction}
 
 Artificial intelligence deals with computational tasks that are difficult to handle by ordinary computers but easy at the human level, such as image  recognition, decision-making, translation between languages, speech recognition, etc. Some early attempts to generalize these concepts into quantum theory can be found in \cite{Kak,Vlasov,Zak,Altaisky,Zhou, Ludermir,Siomau}.  Artificial intelligence in quantum theory has started attracting reasonable attention in the physics community due to the growing interest in artificial intelligence and machine learning globally \cite{Petruccione,Wittek, Schuld,qml,qai,Neto,Lukin,Mangini,Jia,deepq,power}. From a practical point of view, building quantum neural networks with many neurons is still challenging. The same difficulty is seen in building any quantum computational device with many qubits.  To understand the problem better, recall that qubits can be seen as a generalization of the $0$ and $1$ truth values in classical Boolean logic functions, and the main difference from classical bits stems from the possibility of having superposition states between $|0\rangle$ and $|1\rangle$ basis states. As we shall argue below, having ternary possible basis states composing from $|0\rangle$, $|1\rangle$ and a superposition state $\alpha |0\rangle+\beta|1\rangle$, where $\alpha$ and $\beta$ are arbitrary complex numbers not equal to zero, can lead to a minimum computational cost. This fact may explain why, at the very basic level of the universe, quantum matter particles (Fermions) have spin-$1/2$. Consequently, nature at its most fundamental level operates at the minimum computational cost. \\
 It is a widely accepted fact that increasing the number of possible basis states increases the information capacity a system can possess. This fact justifies the need for quantum devices that operate in the multiple-valued threshold logic scheme.  The higher-dimensional quantum information carriers are known in the literature as qudits, where $d$ is not limited to 2 \cite{qudit,Sanders, Wei,Neeley,molecular}. The main advantage of qudits  over qubits is the logarithmic reduction of separate quantum systems needed to span the quantum memory, and thus provides more computational capabilities compared with qubits under the same number of subspaces in both cases \cite{qudit}. \\ 
From a classical computing perspective, the multiple-valued logic has been studied in detail, and many devices have been proposed based on multiple-valued logic architectures, such as the full ternary computer "Setun" and resonant-tunneling transistor \cite{Setun,Hurst,tunneling,Pu,Patel,Kim}. The main motivation was to solve the interconnections problem both on-chip and between chips by increasing the information content per connection \cite{Hurst}.  Since most classical computing processes have an analog at the quantum level, it is possible to build multiple-valued quantum logic devices.  A general article on multiple-valued quantum logic could be read from \cite{Perkowski}. \\ In this work, we extend the scope of using multiple-valued logic in the quantum realm in a novel, systematic way by representing each quantum state as a root of unity placed uniquely on the unit circle. More precisely, we introduce the  multi-valued quantum neurons (MVQNs) and discuss the quantum neural networks based on them. Finally, we suggest the use of the orbital angular momentum (OAM) of light in building multi-valued quantum neurons. 

\section{Qubits and  complexity}
For a given number system, the smaller the radix $r$ the larger the number of digits $d$ required  to express a given quantity. The number necessary to represent a range of $N$ is given by $N=r^{d}$  rounded up to the next integer  value if necessary. Adopting  this terminology, the cost or complexity of the system hardware can be
modeled as \cite{Hurst}
\begin{equation}
C=k(r\times d)= k(r \frac{\log N}{\log r})
\end{equation}
where $k$ is constant. Differentiating the last equation  with respect to $r$ gives the minimum cost, which is when  $r=e=2.718$. Since the radix $r$ must be an integer, the minimum cost happens when $r=3$. This proves that using the basis states $|0\rangle$ , $|1\rangle$ and one superposition state between these two basis states leads to a minimum cost. Finally, before we embark on our main objective of this work, the formulation of multiple-valued quantum logic in the space of holomorphic functions, it is worth mentioning that our terminology of the hardware complexity here is of a qualitative signature and is different from the complexity geometry between quantum states mentioned, for example in \cite{Susskind}.
\section{Multiple-valued threshold quantum logic}
The quantum logic of Birkhoff and  Von Neumann refers to the orthocomplemented lattice of closed linear subspaces of a Hilbert space of quantum states \cite{neumann}. The conjugation is given by the intersection of two linear subspaces, and the negation is given by forming an orthogonal complement \footnote{Let $V$ be a vector space over a field $F$ equipped with a bilinear form $B$. We define $f$ to be left-orthogonal to $g$  and $g$ to be right-orthogonal to $f$ if $B(f,g)=0$}. The main difference between classical and quantum logic stems from the failure of the propositional distributive law in the quantum case, i.e., $P \wedge (Q\lor R)\neq (P\wedge Q) \lor (P\wedge R)$ where the symbols $P,Q$ and $R$ are propositional variables. Thus, disjunction is given by forming the linear space of two linear subspaces. Consequently, the notion of  quantum logic  can be defined straightforwardly. \vskip 5mm
 In Boolean logic, there are two truth values (false and true, or 0 and 1). Using more abstract language, these values are elements of the set $K_{2}=\{0,1\}$. In multiple-valued logic , this can be generalized easily to $K_{k}=\{0,1,\dots k-1\}$. However, it is customary in engineering and especially in the context of neural networks to consider the Boolean alphabet to be $E_{2}=\{-1,1\}$ instead of $K_{2}=\{0,1\}$. The generalization of $E_{2}$ to multiple-valued logic can be achieved using $k$ roots of unity in the complex domain \cite{Aizenberg}, that is,
\begin{eqnarray}
	E_{k}=\{1=\epsilon^{0}_{k},\epsilon_{k},\epsilon_{k}^{2},\dots ,\epsilon_{k}^{k-1}\},
\end{eqnarray}
where $\epsilon_{k}^{j}=(e^{i2\pi/k})^{j}=e^{i\phi_{j}}$ and $\phi_{j}=2\pi j/k$ with $j=0,1,\dots k-1$. Therefore, we have $k$ roots of unity positioned on the unit circle. It is important to emphasize that although the absolute value of these elements is one, they  have different arguments and thus can be uniquely determined on the unit circle as shown in Figure 1 for  $k=5, j \in \{0,4\}$.  

\begin{figure}
\centering
\includegraphics[width=6cm]{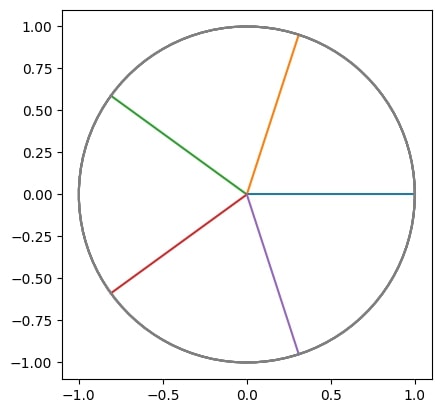}
\caption{The roots of unity for $\epsilon_{5}^{j}$ with $j\in \{0,4\}$. As we will see below, such roots appear in the study of the spin-2 case. }
\end{figure}
\vskip 5mm To connect the above procedure with quantum systems, it is more convenient to use the Bargmann representation of quantum mechanics which maps the quantum states in the vector Hilbert space into analytic functions in the Bargmann space.  In this representation, the bosonic creation and annihilation operators are given by multiplying or taking the derivative  $(\hat{a}^{\dagger}=z, \hat{a}=\frac{\partial}{\partial z})$ acted on a convergent normalized monomial in the complex plane, i.e. $\{\frac{z^{n}}{\sqrt{\hbar^{n} n!}}\}$. This was originally proposed by Fock and developed systematically by Bargmann and Segal\cite{Hall}. More concretely, we define the Segal-Bargmann spaces $\mathcal{H}L^{2}(\mathbb{C}^{d},\mu_{\hbar})$ as spaces of holomorphic functions with Gaussian integration measure $\mu_{\hbar}=(\pi)^{-d}e^{-|z|^{2}/\hbar}$ and  inner product of the form \cite{Hall,almasri1}
\begin{eqnarray}
	\langle f|g\rangle=(\pi \hbar)^{-d}\int_{\mathbb{C}^{d}}\overline{f}(z)\;g (z)\;e^{-|z|^{2}/\hbar}dz, 
\end{eqnarray}
Where $|z|^{2}=|z_{1}|^{2}+\dots +|z_{d}|^{2}$ and $dz$ is the $2d$-dimensional Lebesgue measure on $\mathbb{C}^{d}$. On these spaces, we define the orthonormality condition with $\hbar=d=1$ for simplicity as 
\begin{eqnarray}
	\frac{1}{\pi }\int_{\mathbb{C}}  e^{-|z|^{2}}\; \overline{z}^{n} z^{m}dz =  n!\; \delta_{mn}.  
\end{eqnarray}   
The next step is to apply this representation to quantum mechanical systems. The ideal example to start with is the well-known quantum harmonic oscillator with one degree of freedom, mass $m$, and frequency $\omega$. In this case, the creation and annihilation operators are 
\begin{eqnarray}
	\hat{a}=\sqrt{\frac{m\omega}{2\hbar }}\left(\hat{x}+\frac{i}{m\omega}\hat{p}\right):=\frac{\partial }{\partial z},\\
	\hat{a}^{\dagger}=\sqrt{\frac{m\omega}{2\hbar }}\left(\hat{x}-\frac{i}{m\omega}\hat{p}\right):=z, 
\end{eqnarray}
and the Hamiltonian becomes
\begin{eqnarray}
	H=\hbar \omega \left(z\frac{d}{dz}+\frac{1}{2}\right), 
\end{eqnarray}
with eigenstates $|n\rangle=\frac{z^{n}}{\sqrt{n!}}$.\vskip 5mm 
This allows us to represent important bosonic states such as Fock (number) states, $|n\rangle=\frac{z^{n}}{\sqrt{n!}} $ or coherent states $|\alpha\rangle= e^{-|\alpha|^{2}/2} \sum_{n=0}^{\infty} \frac{(\alpha z)^{n}}{n!}$ in Bargmann representation. The multi-valued logic is represented by the roots of unity.
\begin{eqnarray}
	E_{N}=\{\epsilon^{0}_{N}, \dots \epsilon^{n}_{N}\},
\end{eqnarray}
where $n$ is the energy level index and $N=n+1$. In Figure 2, we show the case of a four-level system with a zero energy ground state for the harmonic oscillator and the corresponding roots of unity.
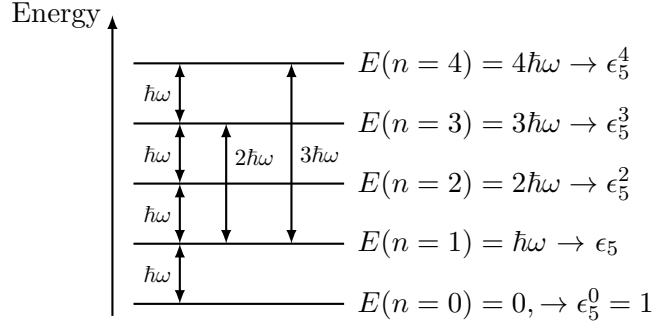
\begin{figure}\label{levels}
	
	\centering
	\begin{tikzpicture}[xscale=2.8,yscale=0.8]
		
		\foreach \n in {0,...,4}{
			\ifthenelse{\n>1}{
				\draw[thick] (0,\n) --++ (1,0) node[right=1pt] {$E(n=\n)=\n \hbar\omega \rightarrow \epsilon_{5}^{\n}$};
			}{\ifthenelse{\n>0}{
					\draw[thick] (0,\n) --++ (1,0) node[right=1pt] {$E(n=\n)=\hbar \omega$ $\rightarrow$ $\epsilon_{5}$};
				}{
					\draw[thick] (0,\n) --++ (1,0) node[right=1pt] {$E(n=\n)=0, \rightarrow  \epsilon_{5}^{0}=1$};
			}}
			\ifthenelse{\n<4}{
				\draw[<->,thick,black!60!black] (0.22,\n) --++ (0,1) node[left,midway,scale=0.8] {$\hbar \omega$};
		}}
		\draw[->,thick] (-0.1,-0.2) --++ (0,5) node[left] {$\mathrm{Energy}$};
		
		\draw[<->,thick,black!60!black] (0.44,1) --++ (0,2) node[above=10pt,right,midway,scale=0.8] {$2\hbar \omega$};
		\draw[<->,thick,black!60!black] (0.75,1) --++ (0,3) node[right,midway,scale=0.8] {$3\hbar \omega$};
	\end{tikzpicture}
	\caption{The first four energy levels for a harmonic oscillator with vanishing ground-state  $E_{0}=0$. In this case, we have the following $k$th roots of unity $E_{5}=\{1,\epsilon_{5},\epsilon_{5}^{2},\epsilon_{5}^{3}, \epsilon_{5}^{4}\}$ for multiple-valued logic.    }
\end{figure}
\vskip 5mm
 For fermions, due to the anticommutation relations between fermionic creation and annihilation operators (being related to Grassmann numbers), it is not possible to apply the Bargmann representation directly to fermions such as electrons. Fortunately, this issue can be solved gently using the Bargmann representation of the Jordan-Schwinger mapping, as discussed, for example, in \cite{almasri2,almasri3}. We can write the operators $J_{i}$ such as spin operators in terms of the holomorphic variables $z$ and $w$  (in natural units $\hbar=1$)  \begin{eqnarray}\label{operators}
	J_{x}=\frac{1}{2}\left(z\frac{\partial}{\partial w}+w\frac{\partial }{\partial z}\right), \\
	J_{y}= -\frac{i }{2}\left(z\frac{\partial}{\partial w}-w\frac{\partial }{\partial z}\right),\\
J_{z}=\frac{1 }{2}\left(z\frac{\partial }{\partial z}-w\frac{\partial }{\partial w}\right), 
\end{eqnarray}
with canonical commutation relation $[J_{i},J_{j}]=i \varepsilon_{ijk} J_{k}$ . 
In this representation, the normalized eigenstates of the spin operators $J^{2}$ and $J_{z}$ are 
\begin{eqnarray}
	|n_{1},n_{2}\rangle=\frac{z^{n_{1}}}{\sqrt{n_{1}!}} \frac{w^{n_{2}}}{\sqrt{n_{2}!}}\equiv f_{n_{1}n_{2}}
\end{eqnarray} 
with normalization condition 
\begin{eqnarray}
	\langle m_{1},m_{2}|n_{1},n_{2}\rangle=\frac{1}{\pi^{2}}\int_{\mathbb{C}^{2}}  \; \overline{f}_{m_{1}m_{2}}f_{n_{1}n_{2}}\; e^{-|z|^{2}-|w|^{2}} dz dw= \delta_{m_{1},n_{1}} \delta_{m_{2},n_{2}}.
\end{eqnarray}

The occupation numbers $n_{1}$ and $n_{2}$ are  connected with the principal angular momentum quantum numbers $j,m$ defined in the basis $|j,m\rangle$ as $J_{z}|j,m\rangle= m|j,m\rangle$ and $J^{2}|j,m\rangle=j(j+1)|j,m\rangle$ through the relations  
	$
	j=\frac{n_{1}+n_{2}}{2}, 
	m=\frac{n_{1}-n_{2}}{2}, 
	$ where $m$ runs from $-j$ to $j$ in integer steps.\vskip 5mm
	
	We define the mapping  from elements of the Bargmann spaces such as $f_{n_{1}n_{2}}$ into a unique set of the roots of unity,  
	\begin{eqnarray}\label{phase}
		f_{n_{1}n_{2}}\rightarrow e^{2\pi i n_{1}/N} e^{2\pi i n_{2}/N} = \epsilon^{n_{1}}_{N} \epsilon^{n_{2}}_{N}
	\end{eqnarray}
where $N=2j+1$ is the number of possible distinct states for a particle or multi-level system with spin number $j$. It is very important to impose the condition $n_{1},n_{2}\in \{0,N-1\}$ in \ref{phase}. \vskip 5mm
\begin{table}
	\caption{\label{demo-table}The mapping  between  the roots of unity and the  state functions for $j=1/2,1,$ and $2$}
	
	\begin{center}
		
		\begin{tabular}{ |c|c|c|c| } 
			\hline
			Roots of Unity 	 & State Function & Quantum Numbers \\
			\hline
			$(\epsilon_{2}^{1})_{z} (\epsilon_{2}^{0})_{w}=\epsilon^{1}_{2}$ & $f_{10}=z$ & $j=1/2, m=1/2$  \\ 
			$(\epsilon_{2}^{0})_{z} (\epsilon_{2}^{1})_{w}=\epsilon^{1}_{2}$	& $f_{01}= w$&  $j=1/2, m=-1/2$ \\ 
			\hline
			$(\epsilon^{1}_{3})_{z} (\epsilon^{1}_{3})_{w}=\epsilon^{2}_{3}$	& $f_{11}=z w$& $j=1,m=0$ \\
			$(\epsilon^{2}_{3})_{z}(\epsilon^{0}_{3})_{w}=\epsilon^{2}_{3}$	& $f_{20}=\frac{z^{2}}{\sqrt{2}}$& $j=1,m=1$ \\
			$(\epsilon^{0}_{3})_{z} (\epsilon^{2}_{3})_{w}=\epsilon^{2}_{3}$	& $f_{02}=\frac{w^{2}}{\sqrt{2}}$ & $j=1,m=-1$\\
			\hline
			$(\epsilon^{4}_{5})_{z} (\epsilon^{0}_{5})_{w}=\epsilon^{4}_{5}$	& $f_{40}=\frac{z^{4}}{\sqrt{24}}$ & $j=2,m=2$\\
			$(\epsilon^{3}_{5})_{z}(\epsilon^{1}_{5})_{w}=\epsilon^{4}_{5}$	& $f_{31}=\frac{z^{3}}{\sqrt{6}}w$ & $j=2,m=1$ \\
			$(\epsilon^{2}_{5})_{z}(\epsilon^{2}_{5})_{w}=\epsilon^{4}_{5}$	& $f_{22}=\frac{z^{2}}{\sqrt{2}}\frac{w^{2}}{\sqrt{2}}$& $j=2, m=0$\\
			$(\epsilon^{1}_{5})_{z}(\epsilon^{3}_{5})_{w}=\epsilon^{4}_{5}$& $f_{13}=z \frac{w^{3}}{\sqrt{6}}$ & $j=2,m=-1$\\
			$(\epsilon^{0}_{5})_{z}(\epsilon^{4}_{5})_{w}=\epsilon^{4}_{5}$	& $f_{04}=\frac{w^{4}}{\sqrt{24}}$& $j=2,m=-2$\\
			\hline
			
		\end{tabular}
	\end{center}
\end{table}
From Table 1, we find that for each spin $j$ the products of the roots of unity are always $\epsilon_{N}^{N-1}$. To distinguish each state with the same spin number $j$ but a different magnetic number $m$, we must take the roots of unity $\epsilon_{N}^{n_{1}}$ in the $z$-plane and $\epsilon_{N}^{n_{2}}$ in the $w$-plane separately. The product of $\epsilon_{N}^{n_{1}}$ and $\epsilon_{N}^{n_{2}}$ is always fixed  and assumes the value
  \begin{equation}
\epsilon_{N}^{n_{1}}\epsilon_{N}^{n_{2}}=\epsilon_{N}^{N-1}
\end{equation}  \vskip 5mm

\section{Complex-Valued Quantum Neural Networks}
Complex-valued neural networks have been studied extensively in the past decades as a natural generalization of the real-valued neural networks\cite{Aizenberg},\cite{Moraga}, \cite{xor}, \cite{Igor} and \cite{Hirose}. Throughout this work, we consider complex-valued neural networks with inputs, activation functions, and outputs represented by roots of unity \cite{Igor}.  It was shown that such complex-valued neural networks enjoy high functionality compared with real-valued neural networks with the same number of neurons\cite{Igor}. More interestingly, a single complex-valued neuron is capable of solving the well-known xor-problem\cite{xor}.\\
Let $Q$ be a quantum system with $N$-distinct energy levels $E_{0},E_{1},\dots , E_{N-1}$. We associate each energy level with a unique root of unity as $\epsilon^{0}_{N},\epsilon^{1}_{N},\dots,\epsilon_{N}^{N-1}$. In case of degeneracy, we associate all equal energy levels into one root of unity. For example, if we have three identical energy levels $E_{n}$, then the corresponding root of unity is $\epsilon_{N}^{n}$, where $N$ is always the number of distinct energy levels.
The multi-valued quantum neuron (MVQN) performs a mapping between $n$ inputs and a single output described by a multiple-valued function $f$ of $n$ variables $f(x_{1},x_{2},\dots x_{n})$ with $n+1$ complex-valued weights
\begin{eqnarray}\label{qn1}
	f(x_{1},x_{2},\dots x_{n})= P(\omega_{0}+\omega_{1}x_{1}+\dots +\omega_{n}x_{n}), 
\end{eqnarray}
where $P$ is the discrete activation function of MVQN : 
\begin{eqnarray}\label{qn2}
	P(z)= \epsilon^{n}_{N}, \;\; 2\pi n/N\leq \mathrm{arg}\;  z < 2\pi (n+1) /N ,
\end{eqnarray}
and $\omega_{0}$ is the bias. Since the root of unity which corresponds to the ground state equals 1. Therefore, we may associate the bias with the ground state. \\ The multi-valued quantum neurons presented here is different from other existing quantum neuron models \cite{Petruccione,Wittek, Schuld,qml,qai,Neto,Lukin,Mangini,Jia,deepq,power}. From a technical point of view, the multi-valued quantum neuron is formulated with inputs, outputs, weights and activation functions written in terms of the roots of unity. This construction allows for straightforward generalization to multiple-valued threshold logic. Physically, it is based on the measured energy eigenstates  of the physical system under study. This in turn would enable for training multi-valued quantum neurons for predicting energy landscape of any quantum system under study with reasonable input. 
\subsection{Hebbian Learning }
Let us  have $n$ $d$-dimensional learning samples. This means that  MVQN has $n$ inputs $(x_{1},\dots x_{n})$. Let ${\bf f}$ be a  $d$-dimensional vector column of the desired output values and $x_{1},\dots x_{n}$ to be a $d$-dimensional vector columns of all possible values of inputs $x^{i}_{1},\dots ,x^{i}_{n}$, where $i=1,\dots d$. The Hebbian learning rule states that the weights $\omega_{1}, \omega _{2},\dots ,\omega_{n}$ must be calculated as dot products of ${\bf f}$ and vectors $x_{1},x_{2},\dots x_{n}$ respectively, 
\begin{eqnarray}
	\omega_{j}= \langle {\bf f}|x_{j}\rangle= f_{1}x_{j}^{1}+\dots+ f_{d} x^{d}_{j}
\end{eqnarray} 
where $j=0,\dots ,n$. The normalized Hebbian rule is given by  division over $d$, 
\begin{eqnarray}
	\omega_{j}=\frac{1}{d} \langle {\bf f}|x_{j}\rangle, \; \; j=0,\dots, n 
\end{eqnarray}
\subsection{Error-Correction Learning  }
Typically, learning algorithms are iterative processes that check whether the inputs and outputs are closer to the desired values with a reasonable error or not. This in turn suggests the need for using an error-correction learning rule that determines how the weights must be adjusted to guarantee that the output is very close to the desired result. The error-correction rule for MVQN with weight vectors $W= (\omega_{0},\omega_{1},\dots,\omega_{n})$ and $X=(1,x_{1},\dots, x_{n})$  is
\begin{eqnarray}\label{learning}
W_{r+1}=W_{r}+\frac{\alpha_{r}}{n+1}\delta \overline{X}
\end{eqnarray}
where $\delta$ is the difference between the desired $\epsilon_{N}^{d}$ and actual $\epsilon_{N}^{a}$ roots of unity, i.e. $\delta= \epsilon_{N}^{d}-\epsilon^{a}_{N}$, $ \alpha_{r}$ is the learning rate, and $\overline{X}$ is the complex conjugate of $X$. The component  $x_{0}$ in $X$ is set to unity since it is associated  with the bias $\omega_{0}$.   The  multi-valued quantum neuron training  is equivalent to movement along the unit circle. Since both the output and desired values are positioned on the unit circle, the movement of the output value towards the desired value will always be in the correct direction. The shortest path for this movement is determined in such a way that the error between the output and desired values is minimized. Let $z= \omega_{0}+ \omega_{1}x_{1}+\omega_{2}x_{2}+\dots +\omega_{n}x_{n}$ be the current weighted sum. The updated weighted sum according to the learning rule \ref{learning} is 
\begin{align}
	z^{\prime}= \omega_{0}^{\prime}+ \omega_{1}^{\prime}x_{1}+ \dots +\omega^{\prime}_{n}x_{n}\\ \nonumber= 
	\left(\omega_{0}+ \frac{\alpha_{r}}{n+1} \delta\right)+ \left(\omega_{1}+ \frac{\alpha_{r}}{n+1}\delta \overline{x}_{1}\right)x_{1}+ \dots + \left(\omega_{n}+ \frac{\alpha_{r}}{n+1}\delta \overline{x}_{n}\right)x_{n}= 
	z+ \alpha_{r} \delta 
\end{align}
 \vskip 5mm To summarize, the setup considered in this work is shown in FIG.3. First we measure  the energy levels  of a given quantum system and associate them with unique roots of unity. These roots of unity are the inputs of a complex-valued neural network. The upper index $n$ of the output $\epsilon_{N}^{n}$ gives the desired energy level of the quantum system under study if the training is correct.  Such a complex-valued neural network is capable of learning the patterns of the energy spectrum a quantum system can have. 
\begin{figure}
	\centering
	\includegraphics[width=10cm]{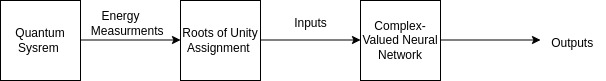}
	\caption{The Complex-Valued quantum neural network architecture.   }
\end{figure}  \section{Applications} The practical execution of the multi-valued quantum neurons can be achieved using equations \ref{qn1} and \ref{qn2}. Now, we show how one could use the multi-valued quantum neurons in determining the maximum and minimum energy eigenvalues for a given quantum system. For the sake of simplicity, we consider a three-level system, in this case, the inputs are elements of $E_{3}=\{1= \epsilon_{3}^{0},\epsilon_{3},\epsilon_{3}^{2}\}$. As mentioned in \cite{Igor}, the definition of the weight vector $W$  determines the operation of learning. For example the weight vector that implements the max-function $f_{\mathrm{max}}= \mathrm{max}(x_{1},x_{2})$ is  $W=\left(-2-4\epsilon_{3}, 4+5\epsilon_{3},4+5\epsilon_{3}\right)$. We may verify this by direct computations of the activation functions and equating the results with the action of the max-function. Since $x_{1,2}\in E_{3}$, the number of all possible distinct inputs is 9, 
\begin{eqnarray}
	x_{1}= \epsilon_{3}^{n_{1}}, x_{2}=\epsilon_{3}^{n_{2}}, 
\end{eqnarray}
with $n_{1}=n_{2}=\{0,1,2\} $. As a sample calculation, if $x_{1}=\epsilon_{3}^{0}$ and $x_{2}=\epsilon_{3}^{2}$. Then, we have 
\begin{eqnarray}
	z=\omega_{0}+ \omega_{1}x_{1}+ \omega_{2}x_{2}=7\epsilon_{3}^{0}+\epsilon_{3}+4\epsilon_{3}^{2}=7+\epsilon_{3}+4\epsilon_{3}^{2},
\end{eqnarray}
\begin{eqnarray}
\mathrm{arg}(z)= 5.7596 \rightarrow P(z)= \epsilon_{3}^{2}=f_{\mathrm{max}}(\epsilon_{3}^{0},\epsilon_{3}^{2}). 
\end{eqnarray}
Analogously , if $x_{1}=\epsilon_{3}^{2}$ and $x_{2}=\epsilon_{3}$. Then, we have
\begin{eqnarray}
z=\omega_{0}+ \omega_{1}x_{1}+ \omega_{2}x_{2}=3+9\epsilon_{3}^{2}
\end{eqnarray}
\begin{eqnarray}
\mathrm{arg}(z)=4.5223  \rightarrow P(z)=\epsilon_{3}^{2}=f_{\mathrm{max}}(\epsilon_{3}^{2},\epsilon_{3}).
\end{eqnarray}
The other remaining 7 possibilities can be calculated in the same manner. The min-function $f_{\mathrm{min}}= \mathrm{min}(x_{1},x_{2})$ can be associated with the weight vector $W=(2+4\epsilon_{3}, 4+5\epsilon_{3},4+5\epsilon_{3})$. If $x_{1}=\epsilon^{0}_{3}$ and $x_{2}=\epsilon_{3}^{2}$, then $f_{\mathrm{min}}(x_{1},x_{2})= \epsilon_{3}^{0}$ since 
\begin{eqnarray}
	z=\omega_{0}+\omega_{1}x_{1}+\omega_{2}x_{2}= 11+9\epsilon_{3}+4\epsilon_{3}^{2}\rightarrow \mathrm{arg}(z)=0.662
\end{eqnarray}
Thus, $P(z)= \epsilon_{3}^{0}=f_{\mathrm{min}}(\epsilon^{0}_{3},\epsilon_{3}^{2})$ as expected. In the same manner, if $x_{1}= \epsilon_{3}^{1}$ and $x_{2}=\epsilon_{3}^{2}$, then $f_{\mathrm{min}}(x_{1},x_{2})= \epsilon_{3}^{1}$ since 
\begin{eqnarray}
	z= \omega_{0}+\omega_{1}x_{1}+\omega_{2}x_{2}=7+8\epsilon_{3}+9\epsilon_{3}^{2} \rightarrow \mathrm{arg}(z)= 3.6652 
\end{eqnarray}
Therefore, $P(z)=\epsilon_{3}^{1}=f_{\mathrm{min}}(\epsilon_{3}^{1},\epsilon_{3}^{2})$. The other remaining possibilities can be computed using similar steps.  
\\The next step is to realize multi-valued quantum neurons experimentally. This would have advantages in building neuromorphic hardware devices that operate with complex inputs, outputs, and weights represented precisely by roots of unity, as we mentioned above. One possible experimental application could be found in the orbital angular momentum of light \cite{Allen}.  This could be done unambiguously by associating the unique root of unity $\epsilon^{n}_{N}$ for each beam with a fixed orbital quantum number $\ell$ in analogy to Table 1 presented in section 3. Other possible systems are molecular spin qudits such as $\mathrm{GdW}_{30}$ molecules\cite{Luis}. It has been found that such molecules can be used as building blocks in future quantum processors due to their tunability, reproducibility, and scalability \cite{Jenkins,Luis1,Luis2}. From a theoretical point of view, one further direction is to consider multilayer complex-valued quantum neural networks and train them to predict the energy configurations of different many-body quantum systems.

			\section*{Acknowledgment}  The author would like to thank the anonymous reviewers for their comments and suggestions. 
			
\end{document}